\documentclass[aps,prc,reprint,superscriptaddress,showpacs,preprintnumbers,floatfix]{revtex4-1}

\usepackage{graphicx}
\usepackage[abs]{overpic} 
\usepackage{color}
\usepackage{longtable}
\usepackage{upgreek}
 \usepackage{placeins}
\usepackage{afterpage}

\usepackage{amsmath}
\usepackage{amssymb}

\usepackage{booktabs}

\begin{document}

\title{Examination of experimental conditions for the production of proton-rich and neutron-rich hypernuclei}

\author{C. Rappold}
\email{c.rappold@gsi.de}
\affiliation{GSI Helmholtz Centre for Heavy Ion Research, Planckstrasse 1, 64291 Darmstadt, Germany}
\affiliation{Universidad de Castilla-La Mancha, Institute of Mathematics applied to Science \& Engineering, Avenida Camillo Jos\'{e} Cela 3, 13071 Ciudad Real, Spain}

\author{J. L\'{o}pez-Fidalgo}
\affiliation{Universidad de Castilla-La Mancha, Institute of Mathematics applied to Science \& Engineering, Avenida Camillo Jos\'{e} Cela 3, 13071 Ciudad Real, Spain}

\date{\today}

\begin{abstract}

After the demonstration of the feasibility of hypernuclear spectroscopy with heavy-ion beams, the HypHI Collaboration will next focus on the study of proton- and neutron-rich hypernuclei. The use of a fragment separator for the production and separation of rare isotope beams is a crucial aspect to producing hypernuclei far from the stability line. Precise spectroscopy of exotic hypernuclei is planned to be carried out at the GSI and later at the FAIR facility with the FRS and Super-FRS fragment separators. A systematic study and an optimization analysis were performed in order to determine optimal experimental conditions for producing hypernuclei with high isospin. The optimal conditions are obtained based on theoretical models for the heavy-ion induced reaction and hypernuclei production.  Experimental efficiencies for the production of exotic secondary beams were also taken into account via Monte Carlo simulations of the fragment separator. The developed methodology is presented to deduce the expected yields of $^8_\Lambda$Be and subsequently other proton-rich and neutron-rich hypernuclei.

\end{abstract}

\pacs{21.80.+a, 25.60.-t, 25.70.Mn}

\maketitle

\section{Introduction}

The interest in strangeness production in nuclear and hadron collisions has been continuously growing during the last decades. In addition to the standard nuclear matter composed by ordinary nucleons, formed by triplets of the two lightest down  and up quarks, the strange (s) quark needs to be considered as well in order to understand the properties of dense matter \cite{cite1,cite2}. The hypernucleus, a bound system of nucleons and hyperons (baryon including at least one s-quark), has demonstrated to be a fundamental tool to study the hyperon-nucleon and hyperon-hyperon interactions \cite{cite3}. At the GSI facility \cite{GSIsite} a research activity on the study of hypernuclei has been carried out since 2006 by the HypHI collaboration \cite{saito_letter_2006}. The next experiments of the HypHI collaboration will then proceed at the future GSI Facility for Antiproton and Ion Research (FAIR). The GSI accelerator facility provides a large variety of ion beams: stable beams from proton to uranium, thanks to the 18-Tm heavy-ion synchrotron (SIS18) \cite{GSI,GSI2}, and exotic beams via the fragment separator (FRS) \cite{FRS,cite6_1}. Ion beams with a kinetic energy up to 2$A$GeV for $A/Z=2$ nuclei can be provided by the current SIS18 synchrotron. The future FAIR facility \cite{FAIRsite} is a substantial expansion of the current GSI accelerator, where additional synchrotron rings of 100~Tm (SIS100) and 300~Tm (SIS300) are planned to be added to the SIS18 of GSI. New experimental apparatuses is under construction for different research programs on nuclear and hadron physics \cite{FAIRdetail}. One of them is the NUSTAR program which foresees the construction of a superconducting fragment separator (Super-FRS) with a magnetic rigidity of 20~Tm to perform powerful in-flight separation of exotic nuclei \cite{geissel2003super}.

The first experiments of the HypHI Collaboration took place in the GSI facility in 2009 and 2010. They succeeded in demonstrating the feasibility of performing a precise spectroscopy of hypernuclei produced in heavy-ion induced reactions \cite{rappold_hypernuclear_2013,PhysRevC.88.041001,Rappold:2015una}. This result was made possible by a novel experimental method, differing from the typical missing mass experiments of mesons or electron beam induced reactions involved at Japan’s National Laboratory for High
Energy Physics (KEK), the Japan Proton Accelerator Research Complex (JPARC), INFN’s Double Annular $\phi$ Factory for Nice Experiments (DA$\Phi$NE), the Thomas Jefferson National Accelerator Facility (JLab), and or the Mainz Microtron (MAMI-C) accelerator \cite{hashimoto_spectroscopy_2006,Botta2012,RepProgPhys78_9}. The first experiment, the Phase 0 experiment, was performed by bombarding a stable $^{12}$C target material with a $^6$Li beam at 2 $A$GeV. The main goal of the experiment was to produce, reconstruct, and identify decay vertexes of $\Lambda$ particle and $^3_\Lambda$H, $^4_\Lambda$H, and $^5_\Lambda$He hypernuclei \cite{saito_letter_2006}. The final results of the data analysis show that the experimental method is viable for the study of hypernuclei \cite{rappold_hypernuclear_2013}. A second experiment with a $^{20}$Ne beam was then performed with similar conditions, and its data analysis is ongoing. 

Several hypernuclear bound states were identified in the same data sample due to the open geometry of the experimental setup, which is a common characteristic of inclusive experiments. In addition, an indication of a possible new bound state has been found: the association of two neutrons and a $\Lambda^0$ hyperon, forming a neutral hypernucleus \cite{PhysRevC.88.041001}. On the other hand, exclusive measurements could provide more precise information on the hypernuclear structure. Consequently, the future HypHI experiments are planned to be performed as exclusive measurements in the fragment separators, FRS or Super-FRS. For instance, recent theoretical calculations disproved the existence of the $^3_\Lambda$n bound state \cite{PhysRevC.51.2905,Gal201493,PhysRevC.89.061302,PhysRevC.92.024325,PhysRevC.91.014003}, which could be either confirmed or denied by a precise exclusive mass measurement. The future phases, namely, Phases 1 and 2 of the HypHI project, focus on the study of exotic hypernuclei toward the proton and neutron drip lines. It will necessarily involve the use of rare-isotope beams, aiming to extend the hypernuclear chart to the proton drip line up to $^{22}_\Lambda$Si hypernuclei and the neutron drip line up to $^{14}_\Lambda$Li hypernuclei. A large charge symmetry breaking effect may be expected in proton- and neutron-rich hypernuclei. A difference between $\Lambda$-proton and $\Lambda$-neutron interactions may induce a shift of the drip-line positions. For this purpose a new experimental apparatus is under development in order to exploit the rare isotope beam provided by the FRS in the actual GSI facility or by the Super-FRS of the future FAIR facility. The Super-FRS fragment separator is then crucial to the future phases of the HypHI project at FAIR for studying proton- and neutron-rich hypernuclei. 

The production of exotic hypernuclei can be influenced by the high isospin of the beam projectile as it will be shown in this article.  The study reported in this article aimed at determining which experimental conditions are necessary for the production of proton-rich or neutron-rich hypernuclei. The feasibility study demonstrating the possibility of operating the Super-FRS at energies around 2 AGeV is presented. The primary beam and target isotopes have to be chosen to obtain the exotic beam of interest at 2~$A$GeV. The selected exotic beam then impinges on the secondary production target to produce the exotic hypernucleus of interest.

 First, the different models and simulations used for this purpose will be presented. The description of the method developed for combining all the information into a multivariate data set follows. This allows us to extract the optimal experimental conditions for any possible hypernucleus of interest.

\section{Simulation processes}

\begin{figure}[!h]
\centering
  \resizebox{!}{35mm}{\includegraphics{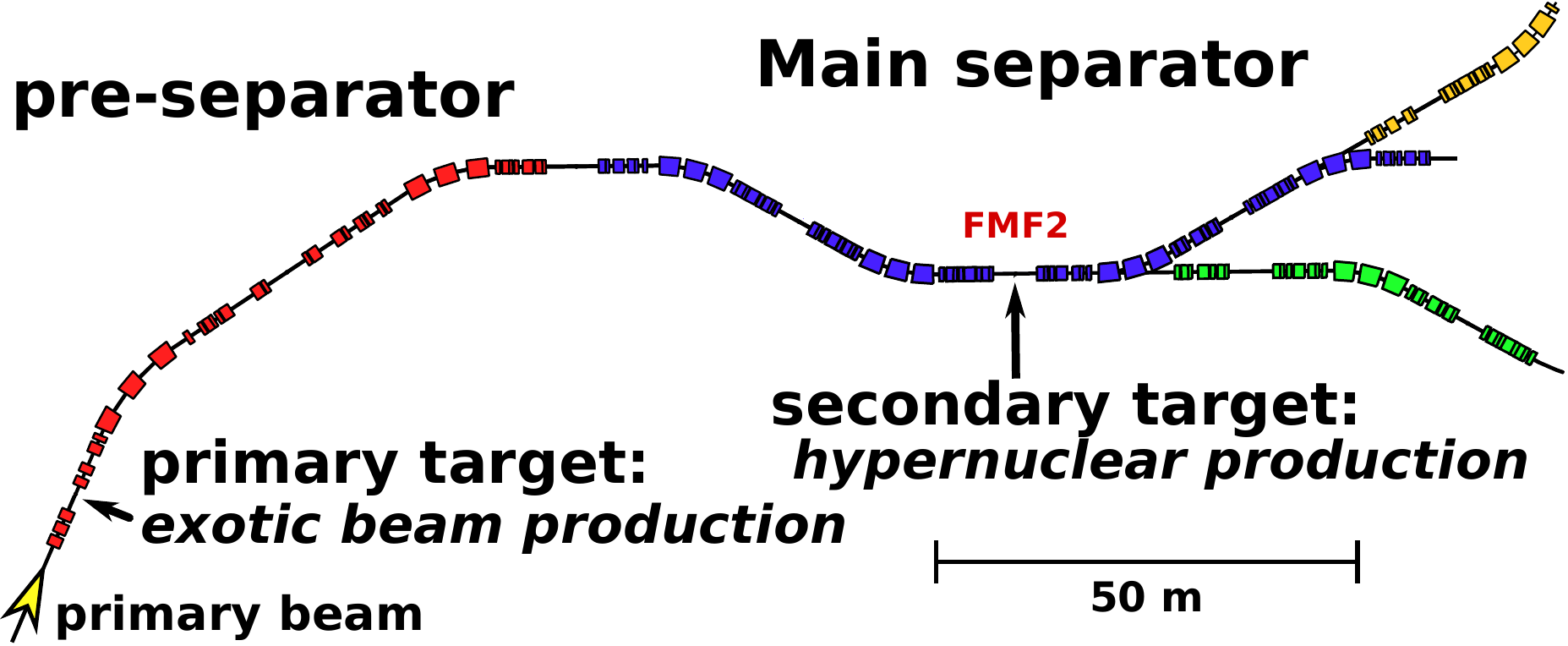}} 
    \caption{ (Color online) Super-FRS fragment separator layout. The primary beam from the synchrotron, represented by the yellow arrow on the beam line, bombards the primary target to produce the exotic beam of interest. The pre-separator of Super-FRS, highlighted in red, is used to select and to produce a high quality exotic beam. The main separator, filled in blue, acts as the high-resolution forward spectrometer. A second target is installed for the production of hypernuclei at the focal plane of the main separator FMF2.}
    \label{fig:SuperFRS_layout}
 \end{figure}

In the forthcoming phases of the HypHI project, within the FRS and Super-FRS fragment separators, the experimental apparatus will focus on exclusive measurements of specific hypernuclei. The production of the desired hypernucleus depends especially on which exotic secondary beam has to be produced and impinges on a secondary target. Figure \ref{fig:SuperFRS_layout} shows the layout of the Super-FRS separator: the primary beam, provided by the synchrotron, bombards the primary target, thus producing exotic fragments. The exotic beam of interest is then selected and purified by the pre-separator, indicated in red in Fig. \ref{fig:SuperFRS_layout}. The exotic beam is delivered to the focal plane, FMF2, of the main separator where a second target for production of hypernuclei is located. The second half of the main separator is used as a high-resolution spectrometer for the decay fragment originated from the mesonic-weak decay of the hypernucleus of interest. The fragment separator will be set to measure a specific momentum range or magnetic rigidity of this decay fragment. This approach will allow a more precise invariant mass measurement of the reconstructed hypernucleus thanks to a momentum resolving power $p/\Delta p=1500$ of the FRS and Super-FRS spectrometers \cite{geissel2003super,winkler2008status}. Therefore, the fragment separator setting and the experimental conditions have to be determined beforehand and a method of data-processing was developed to determine the optimal experimental conditions for the production of a specific exotic hypernucleus.

A first systematic study was performed based on the phenomenological empirical parametrization of
fragmentation cross sections (EPAX) model \cite{PhysRevC.42.2546,epax,PhysRevC.86.014601}. EPAX calculations offer an energy-independent description of the fragmentation cross section at relativistic energy in heavy-ion reactions by means of a universal analytical formula. This phenomenological formula arises from the experimental data sets of the fragmentation reactions of medium- and heavy-ion projectiles. It results in a reasonable estimation of the production cross section of exotic or stable nuclei for a given collision system. 

Numerous beam-target combinations were calculated as an initial data set, establishing the normalized yields of exotic beams per centimeter of production target length. The most interesting beam-target combinations were then preselected. All possible exotic isotopes from hydrogen to scandium have been calculated from all possible combinations of stable isotopes up to $^{40}$Ca. 
Only a subset of isotopes are usable within the FRS and Super-FRS at 2~$A$GeV because of their maximum acceptable magnetic rigidities: 18 and 20~Tm respectively.

\begin{figure}[!hbt]
\centering
   \begin{tabular}{@{}c@{}}
     \resizebox{65mm}{!}{\includegraphics{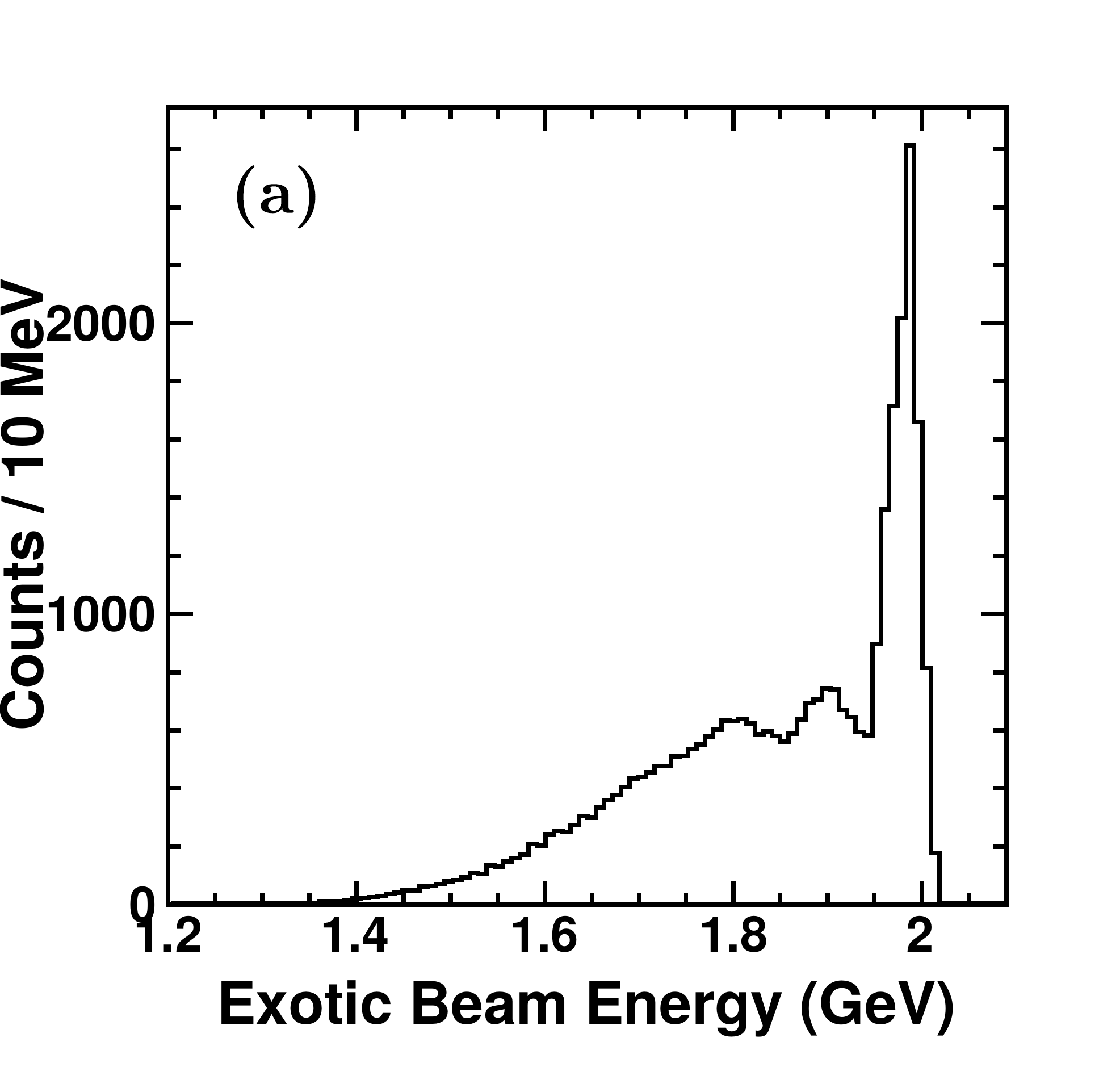} }
         \\[-3mm]
     \resizebox{65mm}{!}{\includegraphics{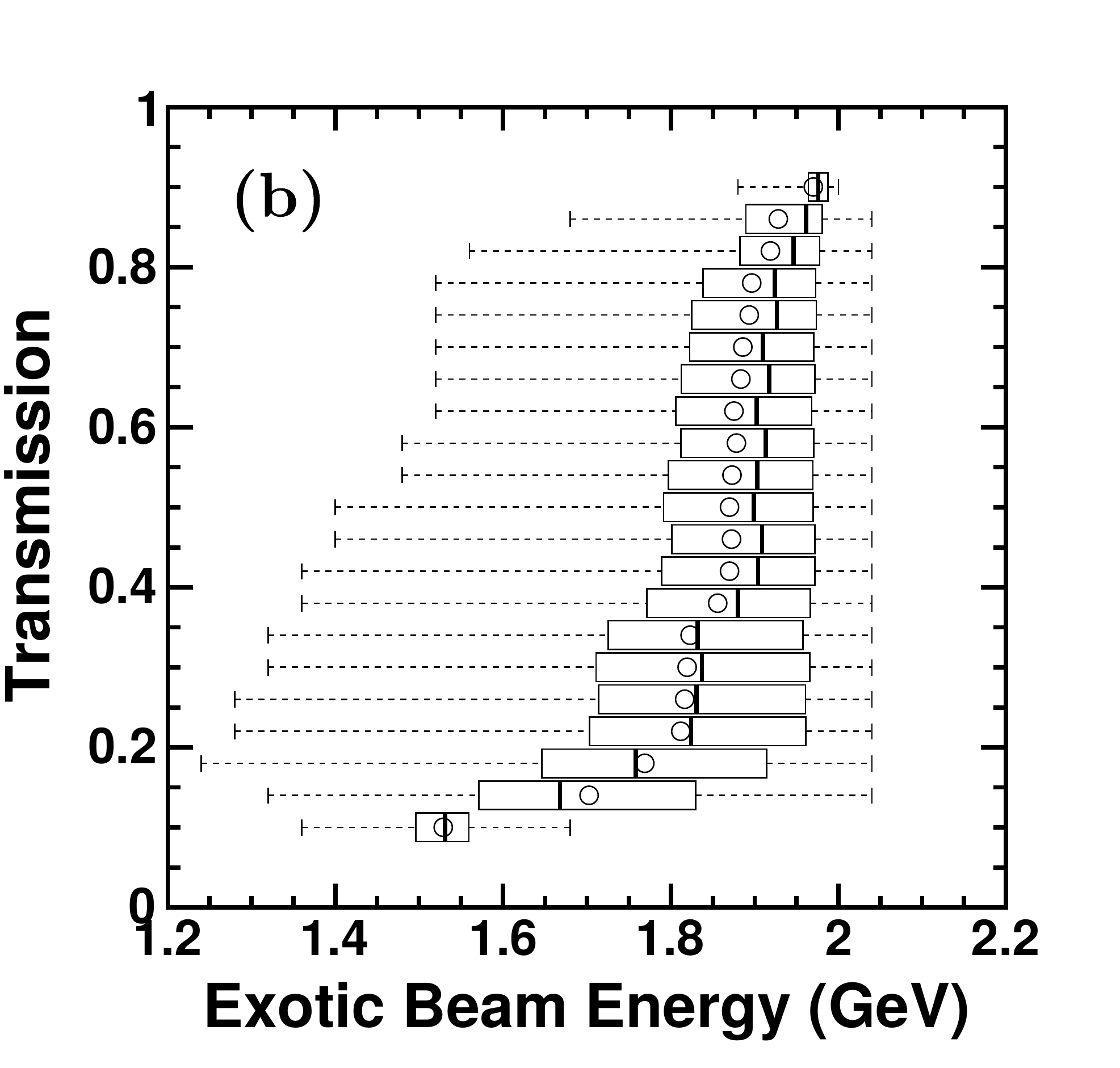}}   
  \end{tabular}
   \caption{Beam kinetic energy distribution and transmission distribution of the fragments propagated in the MOCADI simulations. (a) Projected beam energy distribution. (b) Candle plot of the beam energy distribution as a function of the transmission. For each transmission bin, the box represents the underlying distributions: the bold line represents the quantile at 50\%, the left and right sides of the box are the 75\% and 25\% quantile. The maximum and the minimum of the distributions are represented by the vertical segments of the dotted line. The open circle is the position of the mean value of the distributions.}
   \label{fig:mocadi_par}
\end{figure}

The exotic isotopes yield estimations are integrated into the MOCADI code for the Monte Carlo simulation of the ion-optic transmission \cite{Iwasa2011752,mocadi1}. The availability of a given exotic secondary beam at rare-isotope separator facilities such as the current FRS or future Super-FRS can be then determined. Only the use of the Super-FRS separator was accounted for in the experimental efficiency estimations in the following optimization process. The MOCADI simulations consist of tracing the ions through ion-optical elements in which high-order aberrations of the magnetic field are taken into account. The whole experimental equipment of a fragment separator system is simulated within the MOCADI code. In addition, the nuclear interactions with the material of the detectors are simulated in order to allow direct comparison with high-resolution experimental measurements.

The transmission and yield of each possible exotic secondary beam with the Super-FRS apparatus are estimated with this framework. The secondary beam of interest and other exotic isotopes with a similar magnetic rigidity are transported from the production target location up to the FMF2 experimental area of the Super-FRS, shown in Fig. \ref{fig:SuperFRS_layout}. In the FMF2 experimental area, a secondary production target will be placed for the hypernuclei production. The systematic study then includes the secondary beam yield for each set of beam and target species at several target thicknesses. Within the MOCADI simulations, an optimization procedure was implemented to find and set the optimal parameters of the ion-optical elements of the Super-FRS separator with the aim of obtaining the highest intensity of the secondary beam of interest at FMF2 of the Super-FRS. The Monte Carlo study was performed to achieve 1\% of systematic uncertainties. Figure \ref{fig:mocadi_par} shows the results obtained from the MOCADI simulations. In Fig. \ref{fig:mocadi_par}(a) and \ref{fig:mocadi_par}(b) the distribution of the kinetic energy of the fragment beam and the transmission between the production target and the FMF2 area as a function of the kinetic energy are reported, respectively, for all simulated fragments. A beam energy as close as possible to 2~$A$GeV is necessary for maximizing the hyperon production while keeping a reasonable transmission to the FMF2 experimental area.

\begin{figure}[!htb]
\centering
   \resizebox{65mm}{!}{\includegraphics{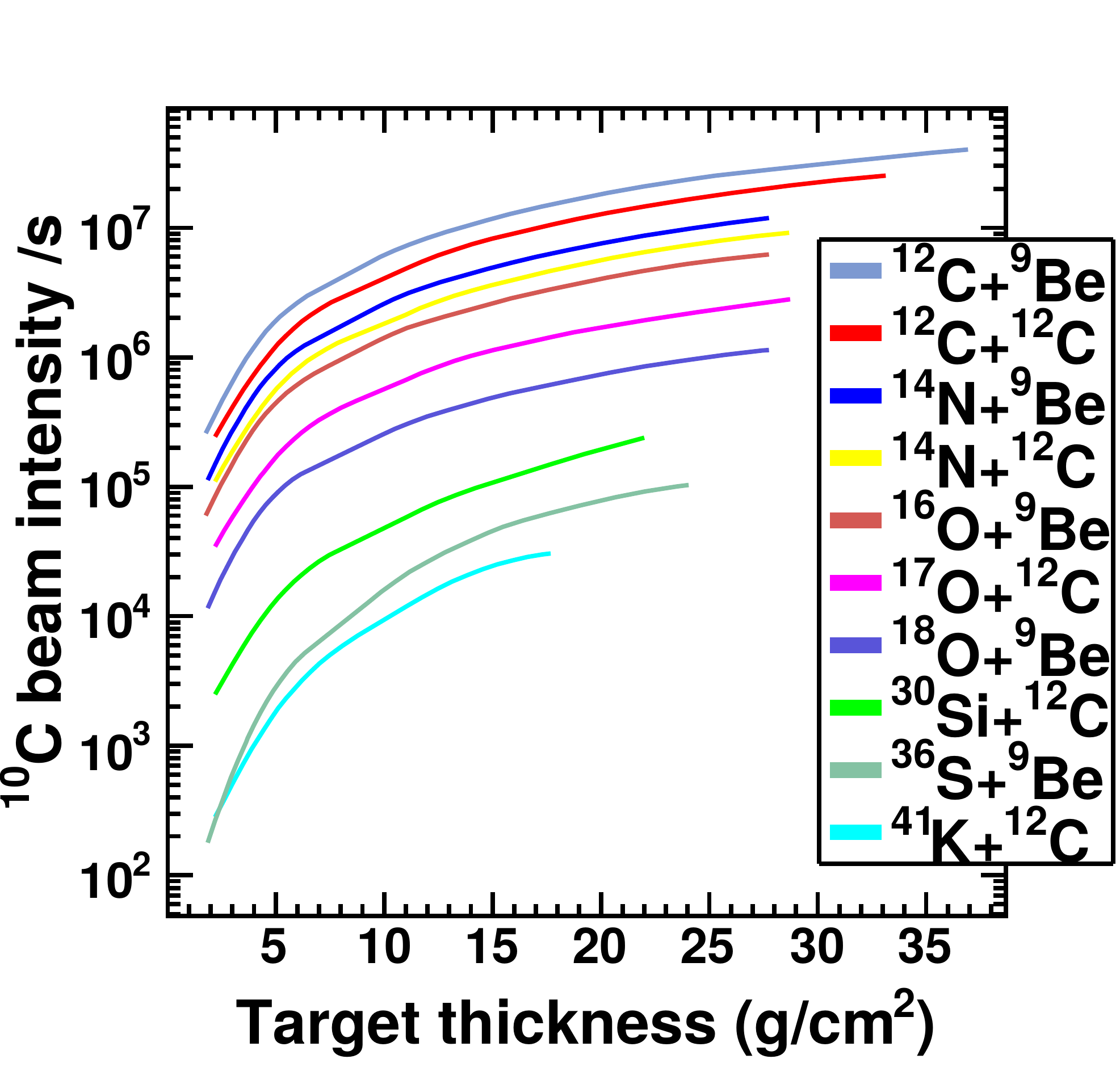}} 
   \caption{ (Color online) Intensity of the secondary beam $^{10}$C at the FMF2 experimental area of the Super-FRS as a function of the production target thickness, the primary beam, and the target isotope. Only ten entries of the results of the MOCADI simulations are displayed out of the whole data set. The intensity of the primary beam was set to $5 \times 10^{9}$ ions/s, which will be available at the Super-FRS. The legend is ordered from highest to lowest intensity.}
   \label{fig:mocadi}
\end{figure}

Additionally, Fig. \ref{fig:mocadi} shows a summary of the $^{10}$C secondary beam production as a function of the target thickness and of the primary-beam and production-target combination. Simulations show that the reaction between (C,~N,~O) beam isotopes and (Be,~C) target isotopes gives the highest $^{10}$C beam intensity up to several million ions/s for a primary beam intensity of $5 \times 10^9$ ions/s. Figure \ref{fig:mocadi} shows the possible optimal case when the target thickness is solely considered for the optimal search. However, other parameters such as the contamination of other produced exotic isotopes, the beam energy, or the secondary reaction for the hypernuclei production have been considered.

\begin{figure*}[!htb]
\centering

   \begin{tabular}{@{}c@{\hskip +3.4mm}c@{}}

   \resizebox{65mm}{!}{ \includegraphics{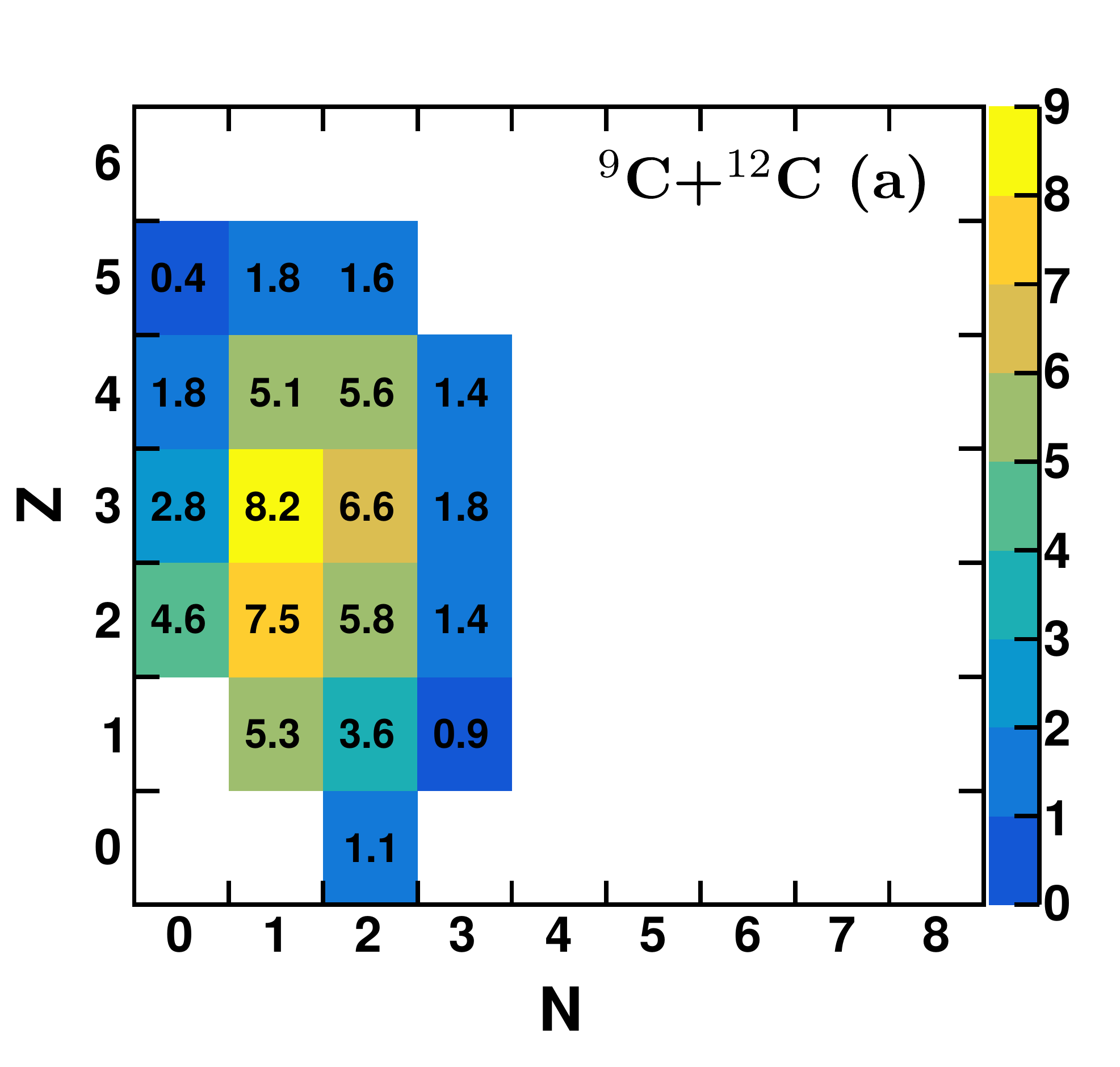} } & 
   \resizebox{65mm}{!}{ \includegraphics{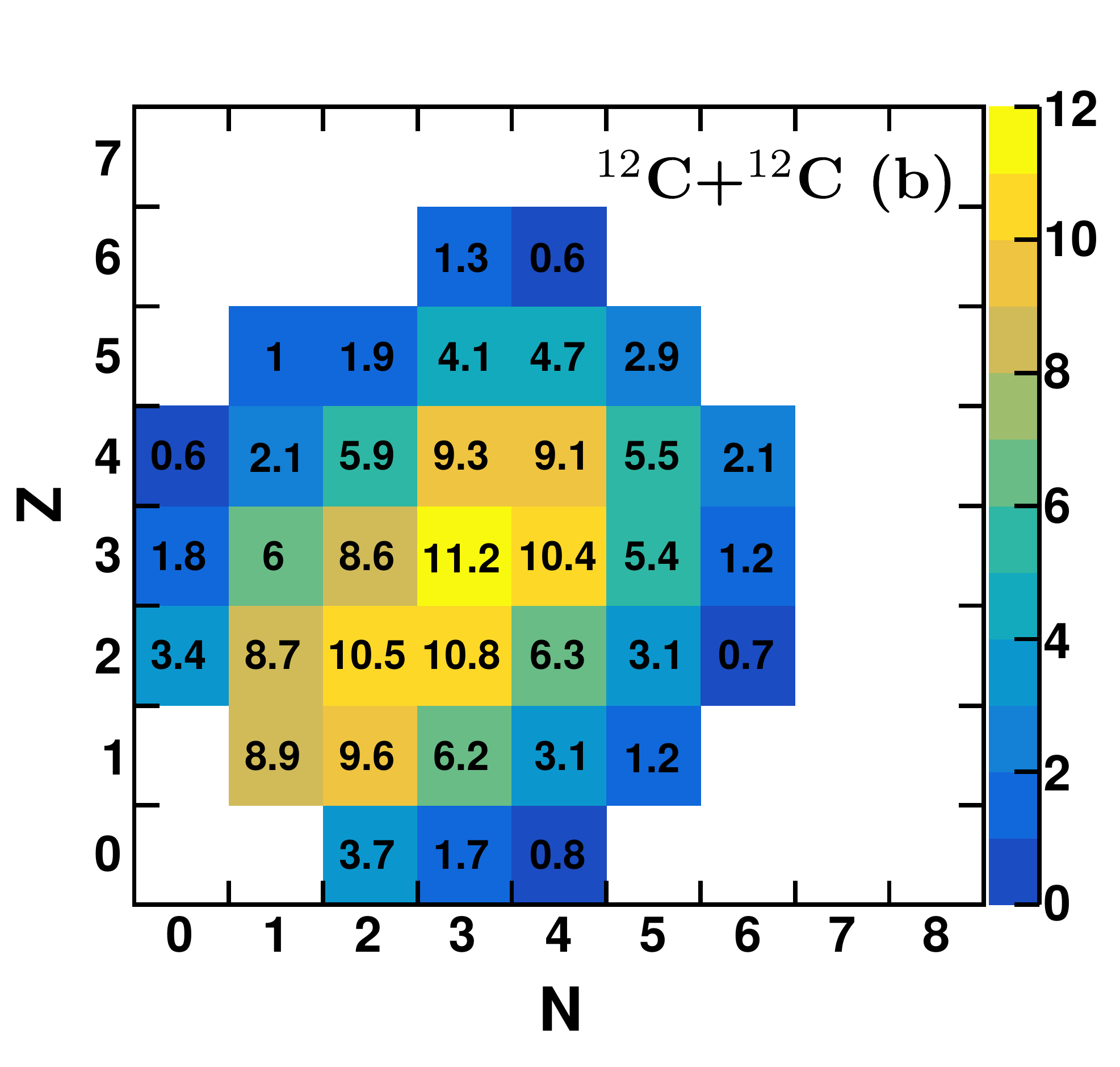} }
\\[-3mm]
   \resizebox{65mm}{!}{ \includegraphics{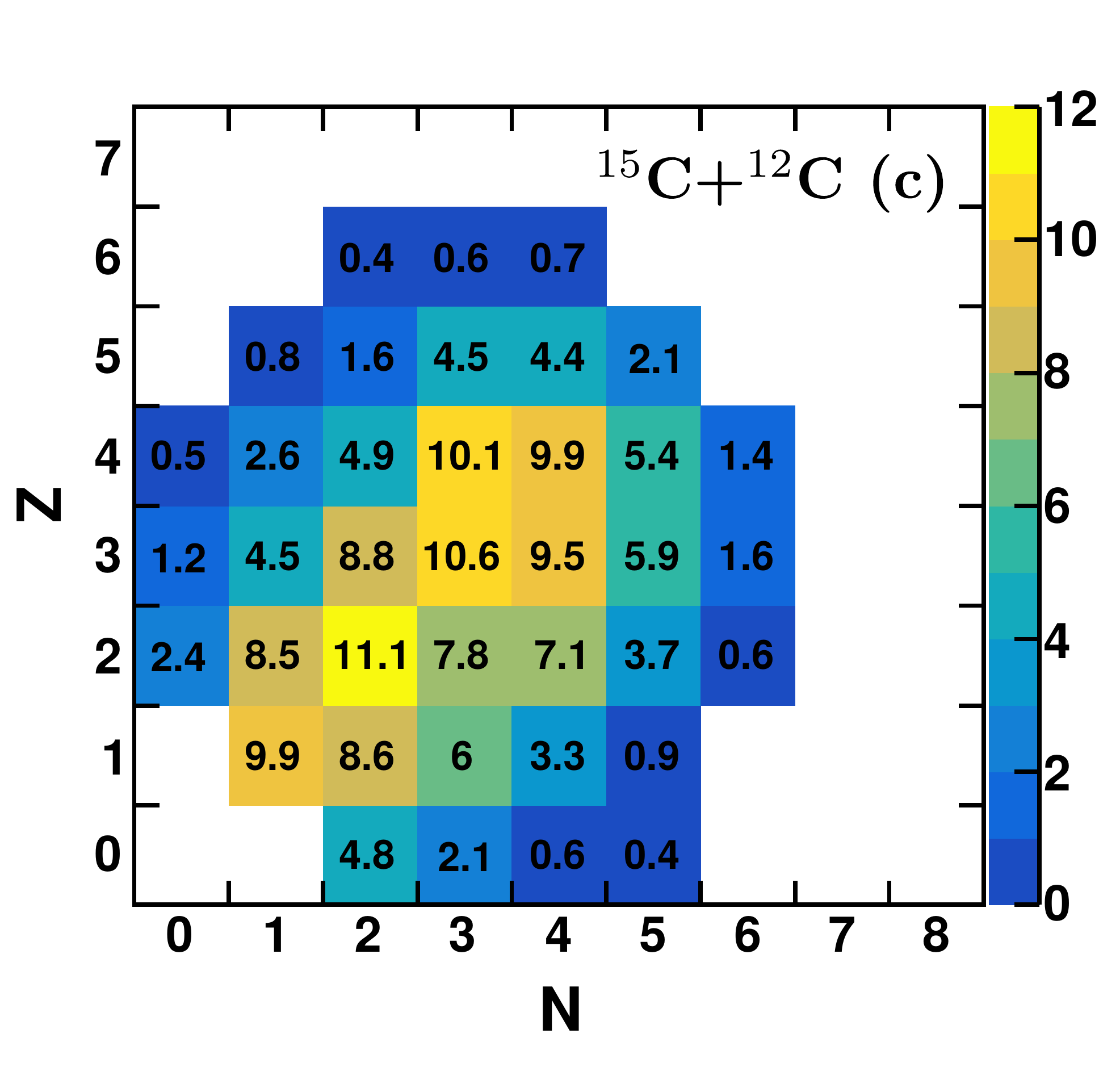} } & 
   \resizebox{65mm}{!}{ \includegraphics{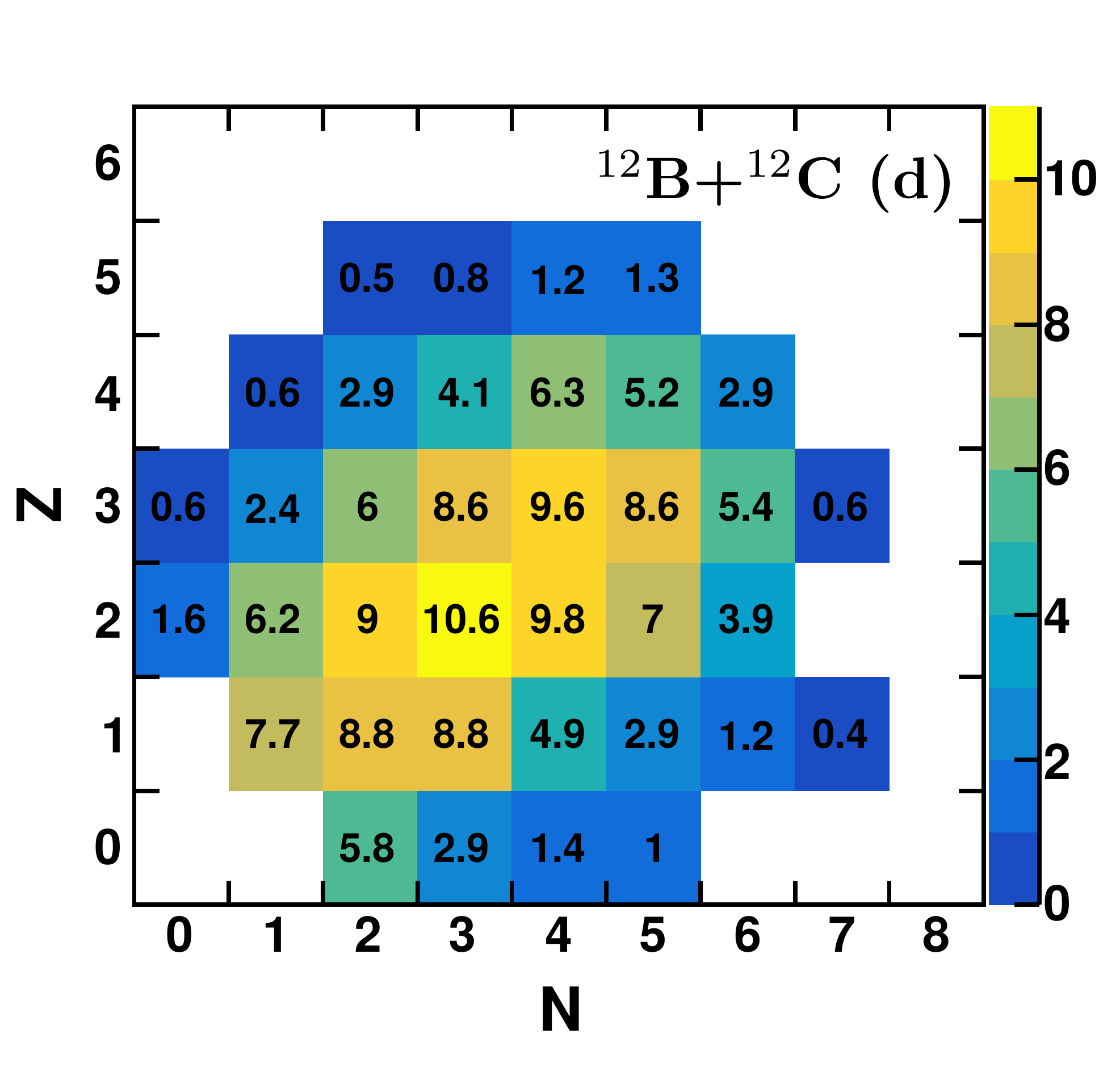} }

   \end{tabular}

  \caption{\label{fig:qgsm_col} (Color online) Production cross section in $\upmu$b of $\Lambda$-hypernuclei from the collision of (a) $^{9}$C+$^{12}$C  at 2~$A$GeV, (b) $^{12}$C+$^{12}$C, (c) $^{15}$C+$^{12}$C, and (d) $^{12}$B+$^{12}$C. The neutron and proton numbers of the $\Lambda$-hypernucleus core are represented by the horizontal and vertical axis respectively. The systematic uncertainty of the production cross sections is estimated to the level of 0.1~$\upmu$b. }
\end{figure*}

The production of the exotic hypernuclei for each secondary beam of interest can then be studied thanks to the theoretical model of Ref. \cite{PhysRevC.88.014611}. It is an hybridization between the transport model DCM-QGSM (Dubna cascade model--quark-gluon string model), which simulates the collision between the beam and the target, and a statistical approach of the Fermi break up model to describe the de-excitation of spectators. The hypernuclei production was investigated for each nucleus-nucleus collision of exotic beam and target species at 2 $A$GeV. For each theoretical calculation a suitable number of events were simulated in order to keep the systematic uncertainty on the hypernuclei production cross sections to the level of 0.1~$\upmu$b. Moreover, the theoretical calculation for $^6$Li+$^{12}$C collisions at 2 $A$GeV was performed and compared with the published experimental results \cite{Rappold:2015una}. The theoretical estimations of the hypernuclei production cross section were compatible with the experiment, validating the calculations of the other colliding systems. Theoretical calculations for the hypernuclei production were carried out according to the exotic beam, from Li to Ne isotopes, colliding on a $^{12}$C or $^{9}$Be target at 2$A$GeV. The obtained results were gathered into a data set, which can be ordered by colliding isotopes or by produced hypernuclei.

Figure \ref{fig:qgsm_col} shows the $\Lambda$-hypernuclei production cross section in $\upmu$b as a function of the neutron and proton numbers of the core for a proton-rich $^{9}$C and a neutron-rich $^{15}$C secondary beam on a $^{12}$C target. A clear difference is observed with respect to the case of the $^{12}$C+$^{12}$C collision, where the exotic carbon beams enhance exotic hypernuclei production from 1.2 to 3 times. Concerning the production of neutron-rich hypernuclei, higher increase of the production cross section is observed in the $^{12}$B exotic beam compared to the $^{12}$C beam as shown in Fig. \ref{fig:qgsm_col}. Choosing the proton-rich exotic beam clearly favors the production of proton-rich~hypernuclei, and reciprocally a neutron-rich beam to produce a neutron-rich~hypernuclei.

\begin{figure*}[!ht]
\centering
    \begin{tabular}{@{}c@{}c@{}}

   \resizebox{!}{65mm}{ \includegraphics{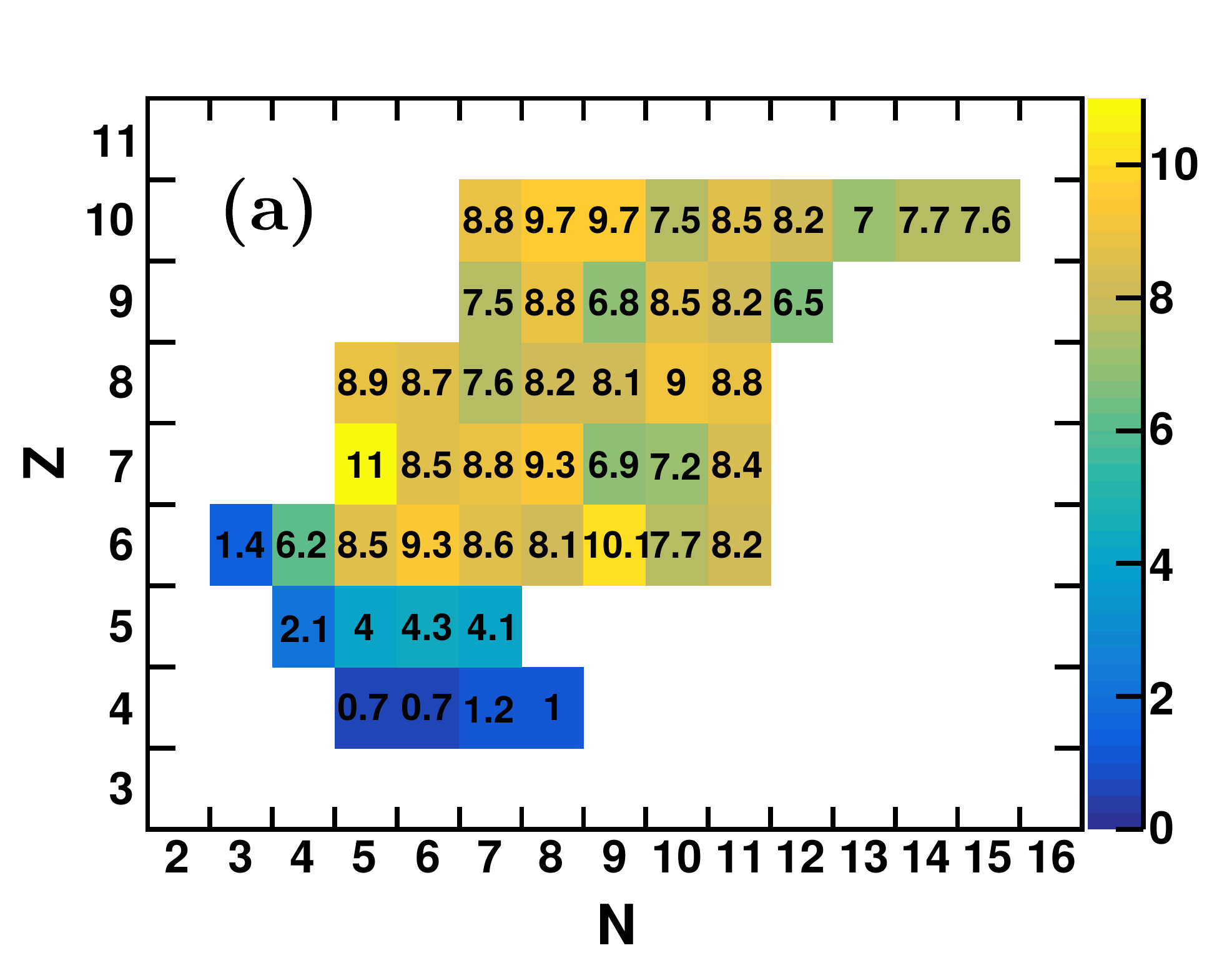} }  & 
   \resizebox{!}{65mm}{ \includegraphics{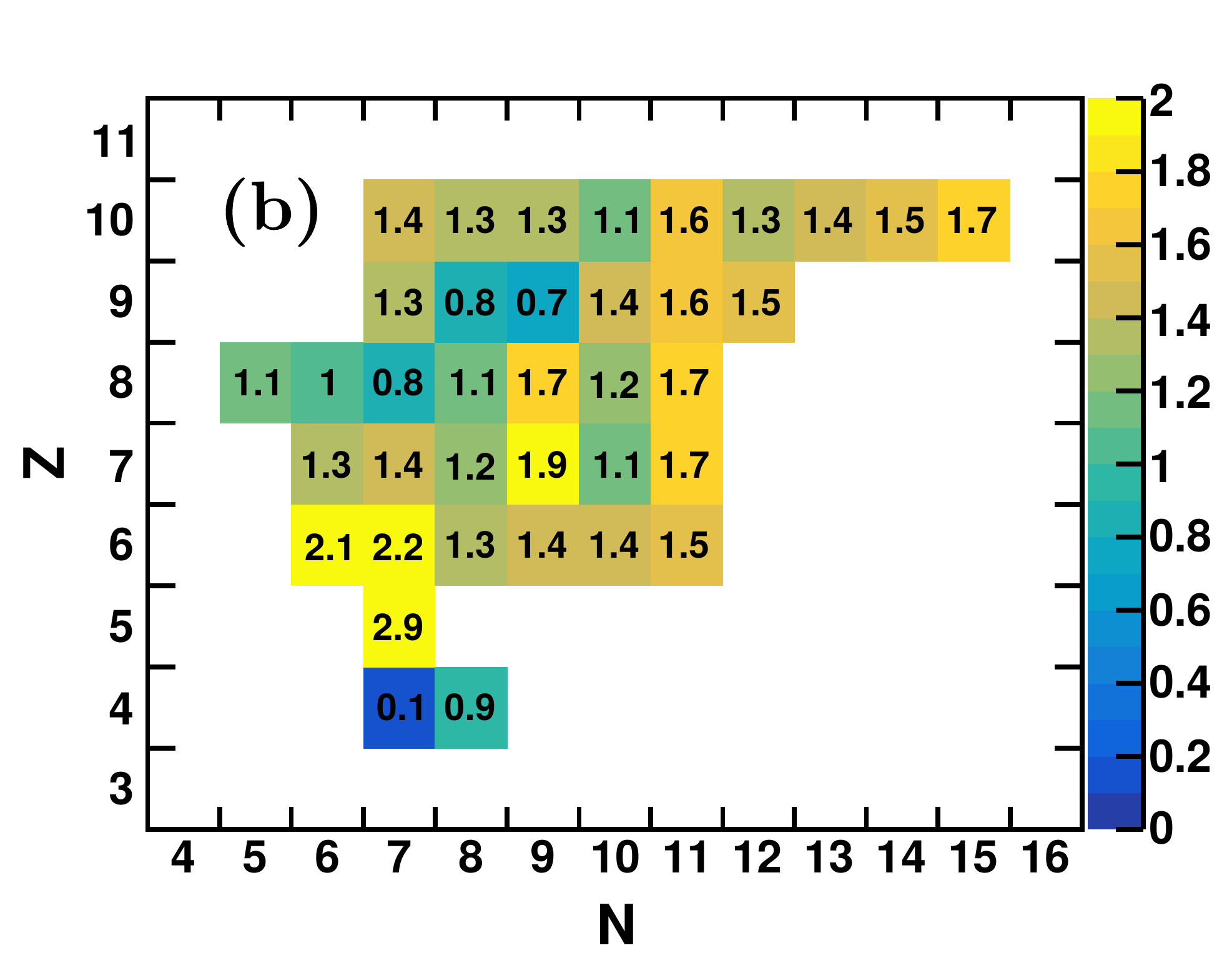} }

   \end{tabular}

  \caption{\label{fig:qgsm} (Color online) Production cross section (in $\upmu$b) of (a) $^8_\Lambda$Be  and (b) $^{11}_\Lambda$Be hypernuclei, according to the collision of different exotic secondary beams $^{Z+N}$Z on a $^{12}$C target at 2~$A$GeV. The systematic uncertainty of the production cross sections is estimated to the level of 0.1~$\upmu$b.}

\end{figure*}

When the results are ordered by the produced hypernucleus, Fig. \ref{fig:qgsm} shows the production cross section of proton-rich hypernucleus $^8_\Lambda$Be and neutron-rich hypernuclei $^{11}_\Lambda$Be at 2$A$GeV depending on the neutron and proton numbers of the exotic beam reacting on a $^{12}$C target. The secondary beam isotope that maximizes the production of each exotic hypernucleus can be then identified. However, this maximum is not necessarily the optimal since the production of the exotic beam has to be also considered. Moreover, the hypernuclei production cross sections need to be adjusted since the kinetic energy of the exotic beam may vary from 2$A$GeV. The parametrization \cite{Faldt_paraLambda_1997} used to fit the world data set of total production cross sections of pp~$\to$~pK$^+\Lambda$ \cite{epja_COSY_2010} can be employed to scale the theoretical calculations. The parametrization from \cite{epja_COSY_2010} was used. The value at 2~$A$GeV was used as a normalization factor to obtain the scaling function in this study. For instance, the hypernuclei production cross section is then reduced by 73\% or 46\% if the energy of the exotic beam is decreased to 1.9~$A$GeV or 1.8~$A$GeV respectively.  

A multivariate data set was then created in order to find the optimal set of experimental parameters to maximize the production of a hypernucleus of interest. It gathers the results of those theoretical calculations and Monte Carlo simulations. This parameter set is defined by the primary beam, the production target, its thickness, and the exotic secondary beam that are optimal to produce the hypernucleus of interest. A generic approach was developed, so that the optimal experimental setup can be determined for each hypernucleus of interest.

\section{Multivariate analysis}

It is merely impossible to simply plot the multivariate data set and estimate the best case by compiling all the possible combinations and permutations of the isotopes species. An optimization procedure was used to find the optimal case. This article focuses first on the production of the proton-rich hypernucleus $^8_\Lambda$Be, yet other hypernuclei are considered afterwards since the optimization procedure does not depend on the hypernuclear species.

During the multivariate analysis, the production of the hypernucleus of interest is taken into account: all possible secondary beams on Beryllium or Carbon targets were considered in the theoretical calculations of the hypernuclei production. The secondary beam selected by the procedure is used to find the optimal conditions of the Super-FRS, which allow us to calculate the hypernuclear yield per second for a 4-centimeter secondary production target. The secondary target thickness was selected to match the experimental condition of the previous HypHI experiments. This hypernuclear yield estimation includes the experimental efficiency of the ion-optic transmission and the exotic beam intensity obtained from a primary beam of $5\times 10^9$ ions/s. Additionally, the simple case of using a stable beam for the hypernuclei production is also included in the data set. An intensity of $10^7$ ions/s was selected in those cases to estimate the hypernuclear yield per second.

In an optimization problem, a cost function has to be defined between the different variables and parameters in order to find the optimal set which maximizes or minimizes this cost function. The cost function was defined as follow for our multivariate data set: 
\begin{equation} 
\label{eq:cost}
F_{\alpha,\beta}(\mathit{C}, \mathit{E}, \mathit{T}, \mathit{I}) = \alpha\, \mathit{C} + \beta\,\mathit{E} - \gamma\,\mathit{T} + \delta \,\mathit{I}, 
\end{equation} 
in which the variables $\mathit{C}$,  $\mathit{E}$, $\mathit{T}$, and $\mathit{I}$ refer to the hypernuclear yield, secondary beam energy, production target thickness, and intensity of the secondary beam of interest, respectively. The parameters $\alpha$, $\beta$, $\gamma$ and $\delta$ defined in the cost function $F$ are the weight coefficients that connect the different variables. The weight coefficient $\delta$ for the intensity parameter was fixed to 1/2 for achieving the numerical stability of the convergence of the cost function to its optimum. Besides, the sum of the squared weights is set to 1, $\alpha^2+\beta^2+\gamma^2+\delta^2=1$, resulting in the corresponding coefficient $\gamma$ being equal to $\sqrt{3/4-\alpha^2-\beta^2}$ and in a constraint for $\alpha$ and $\beta$ being within a circle of radius $\sqrt{3/4}$. In the cost function, the different parameter distributions were normalized to be within a [0, 1] interval in order to keep the weight coefficients within the unit interval. The parameters $\alpha$ and $\beta$ are set arbitrarily depending on the weight to be associated to each variable of the data set. This cost function is built intentionally to maximize the production of the hypernucleus of interest, such as $^8_\Lambda$Be, by obtaining an optimal energy and intensity of the secondary beam while minimizing the thickness of the production target.

\begin{figure}[htb]
\centering
   \resizebox{65mm}{!}{\includegraphics{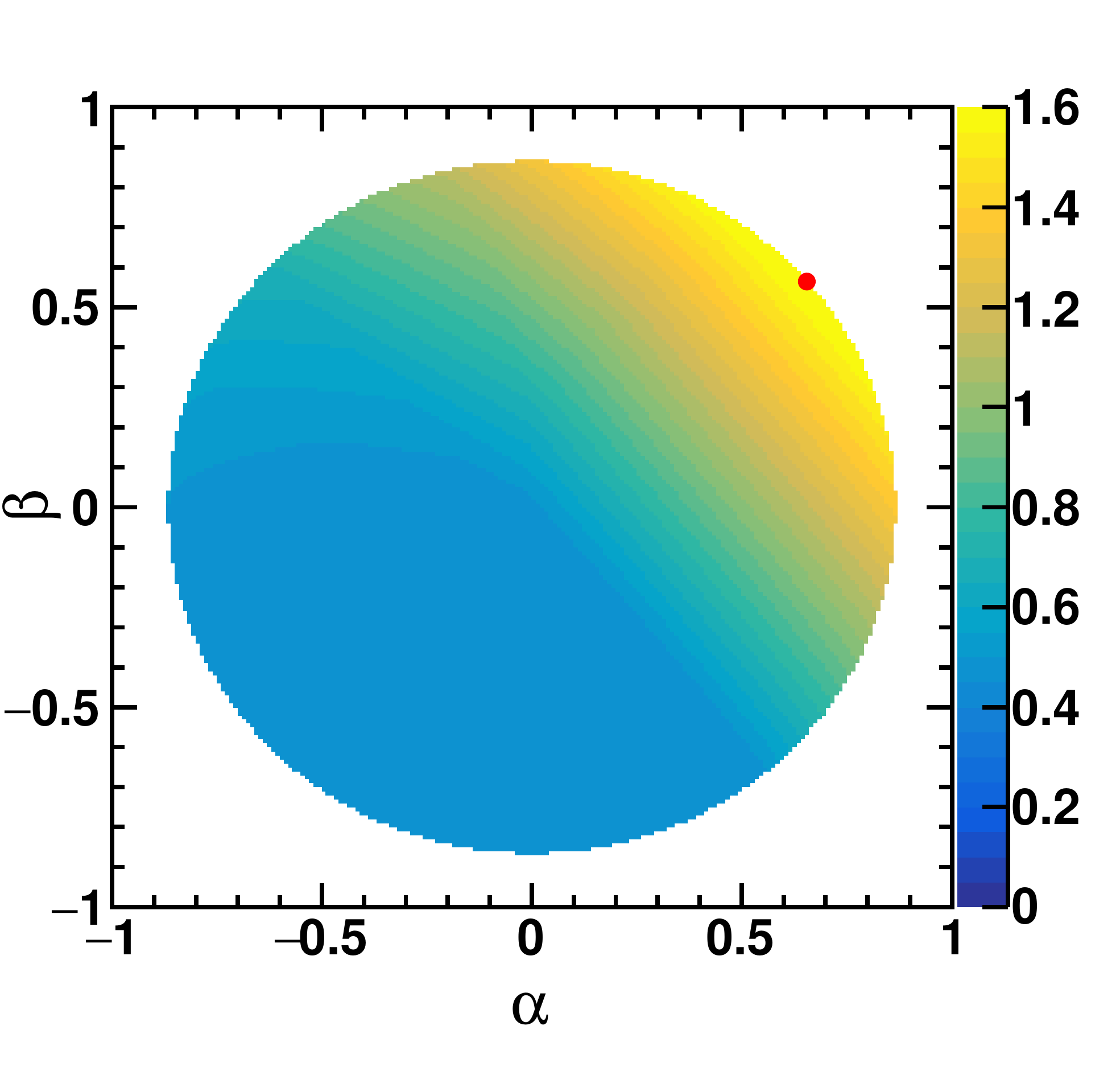}} 
   \caption{   \label{fig:ana_cost} (Color online) The evolution of the cost function maximum as a function of the $\alpha$ and $\beta$ weight coefficients for the multivariate data set optimization of the studied case of $^8_\Lambda$Be. The red dot represents the position of the maximum of the maxima within the $\alpha$-$\beta$ space.}
\end{figure}

\begin{figure*}[htb]
   \begin{tabular}{@{\hskip -4.5mm}c@{\hskip -5.5mm}c@{\hskip -.5mm}c@{}}

   \resizebox{65mm}{!}{ \includegraphics{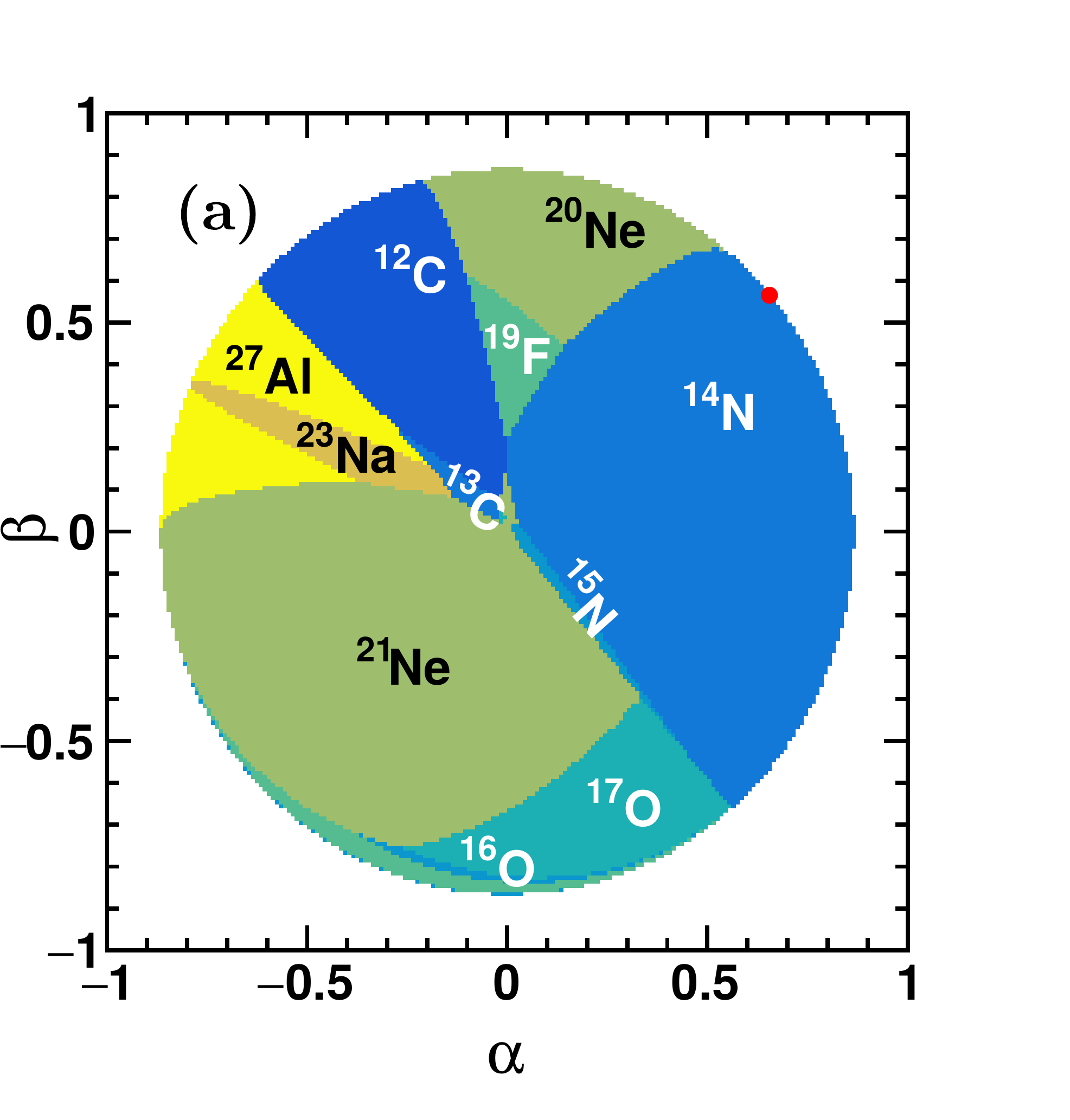} } &
   \resizebox{65mm}{!}{ \includegraphics{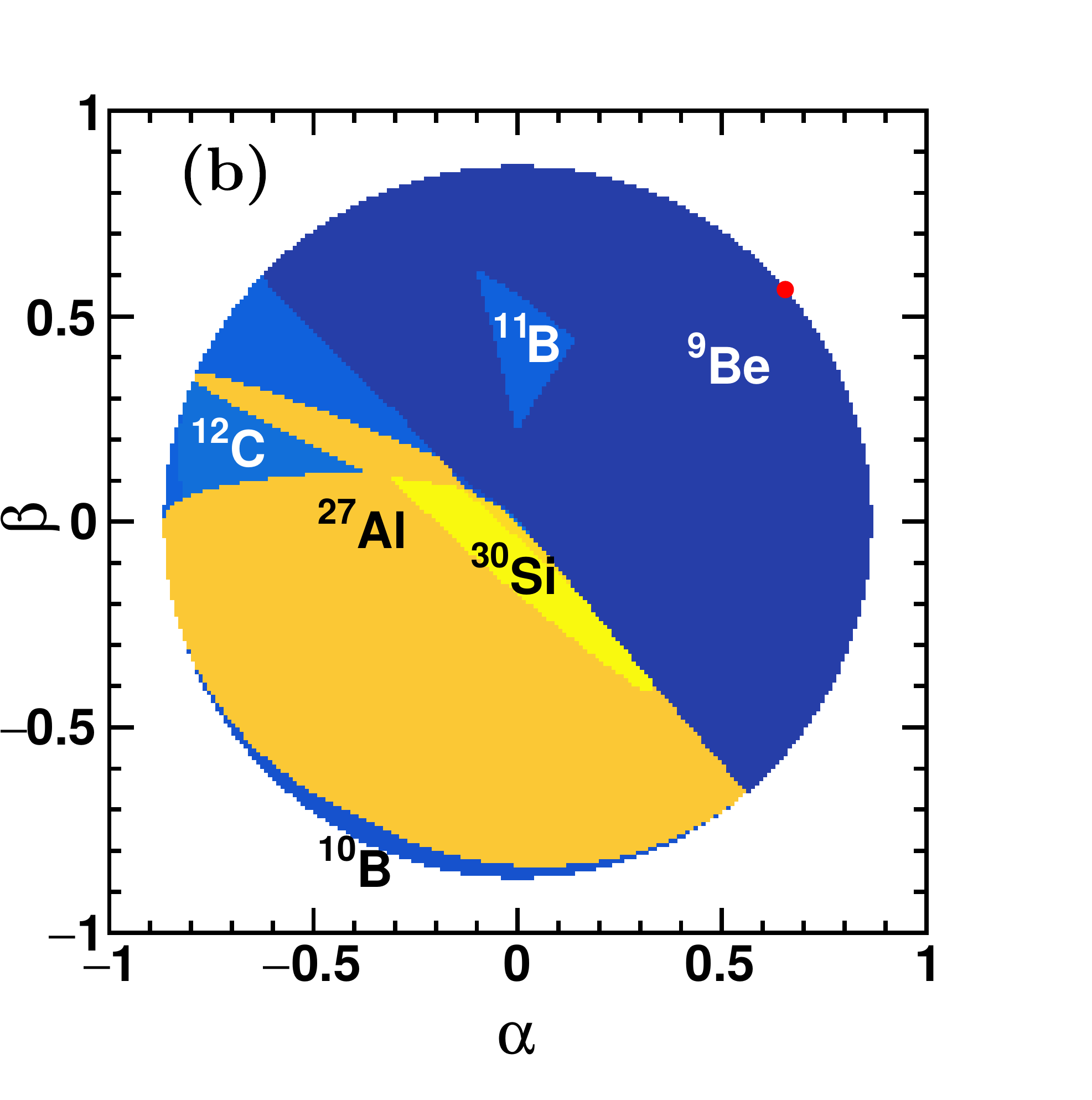} } & 
   \resizebox{65mm}{!}{ \includegraphics{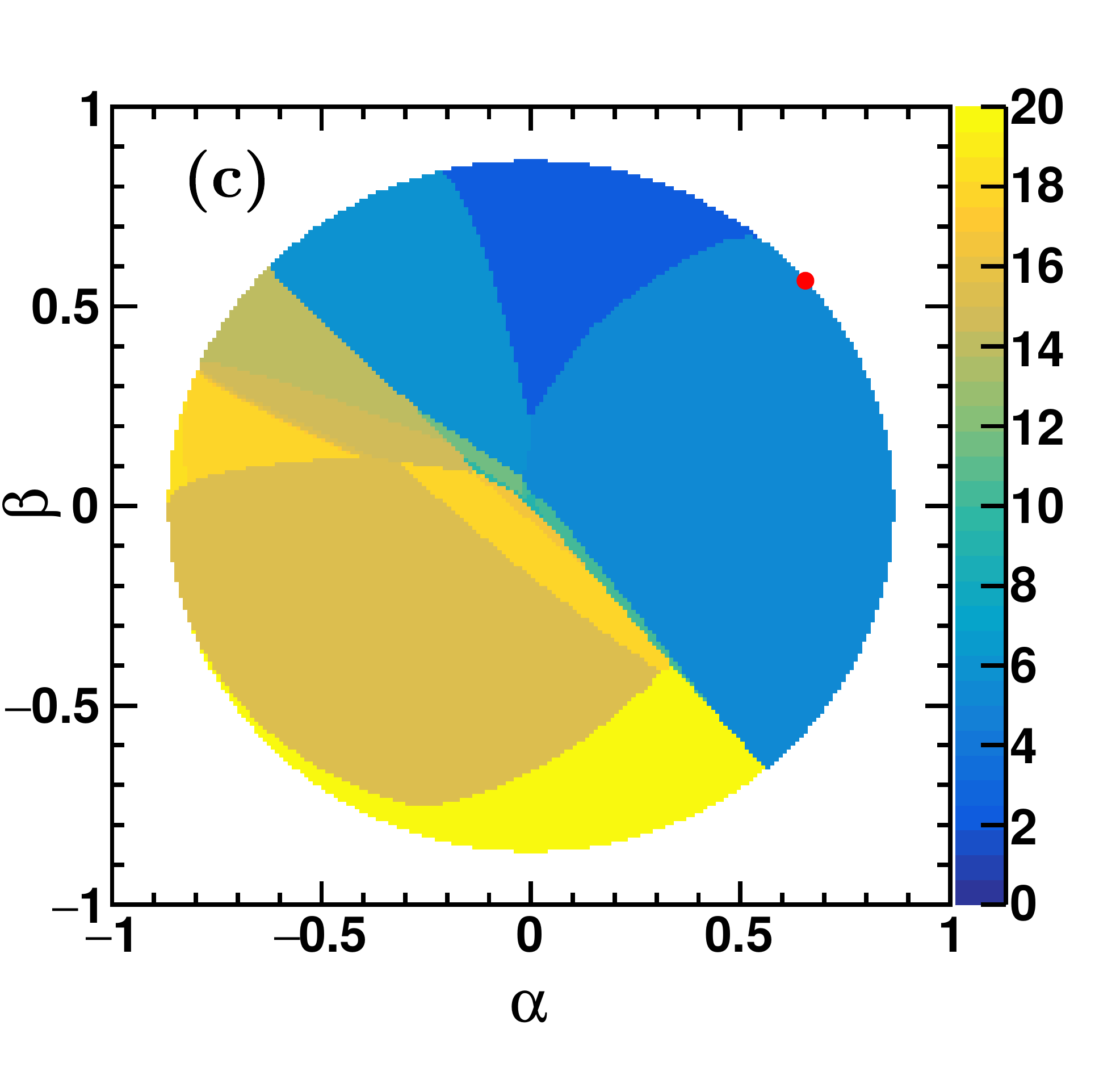} }
\\[-3mm]
   \resizebox{65mm}{!}{ \includegraphics{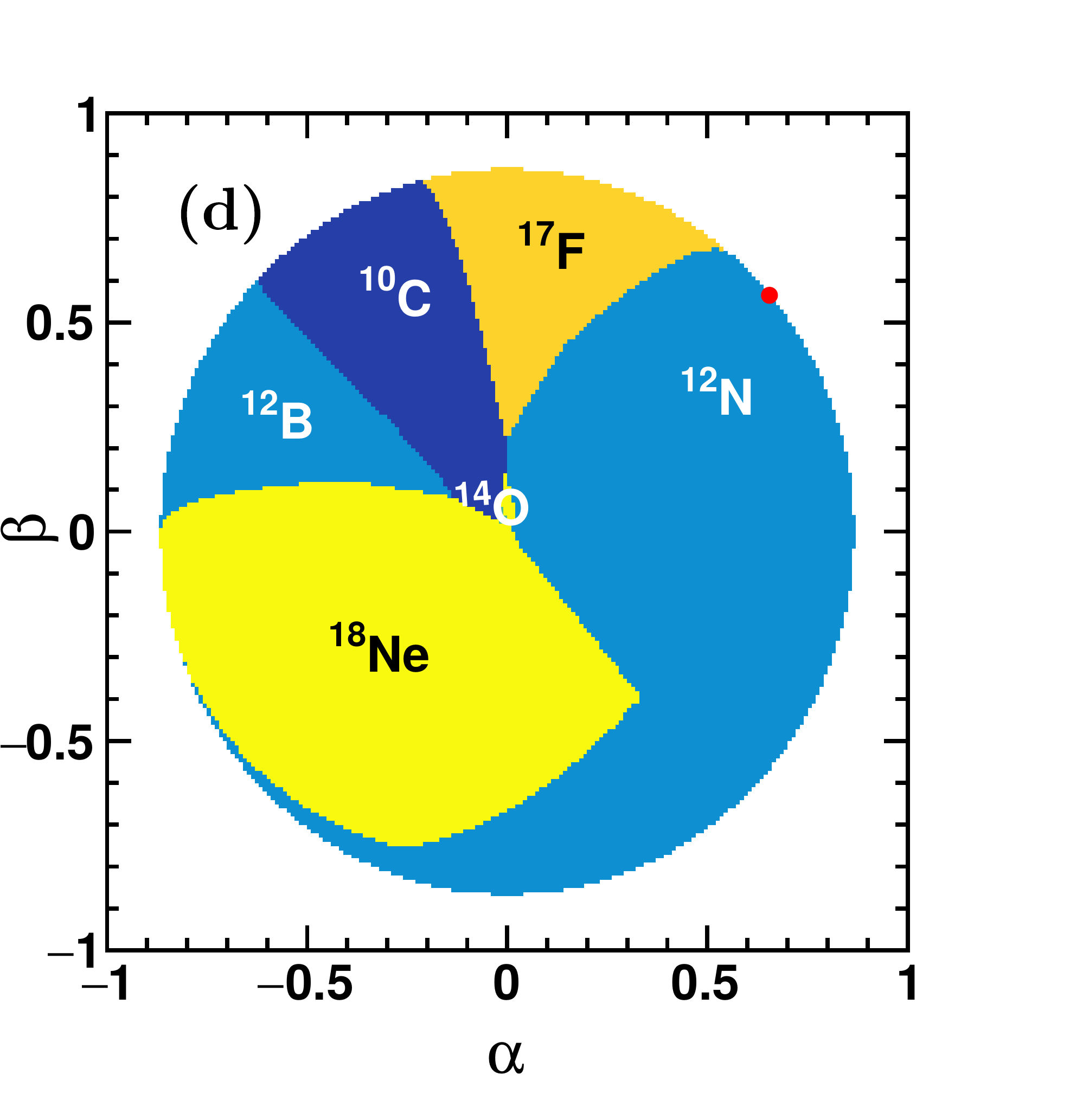} } &
   \resizebox{65mm}{!}{ \includegraphics{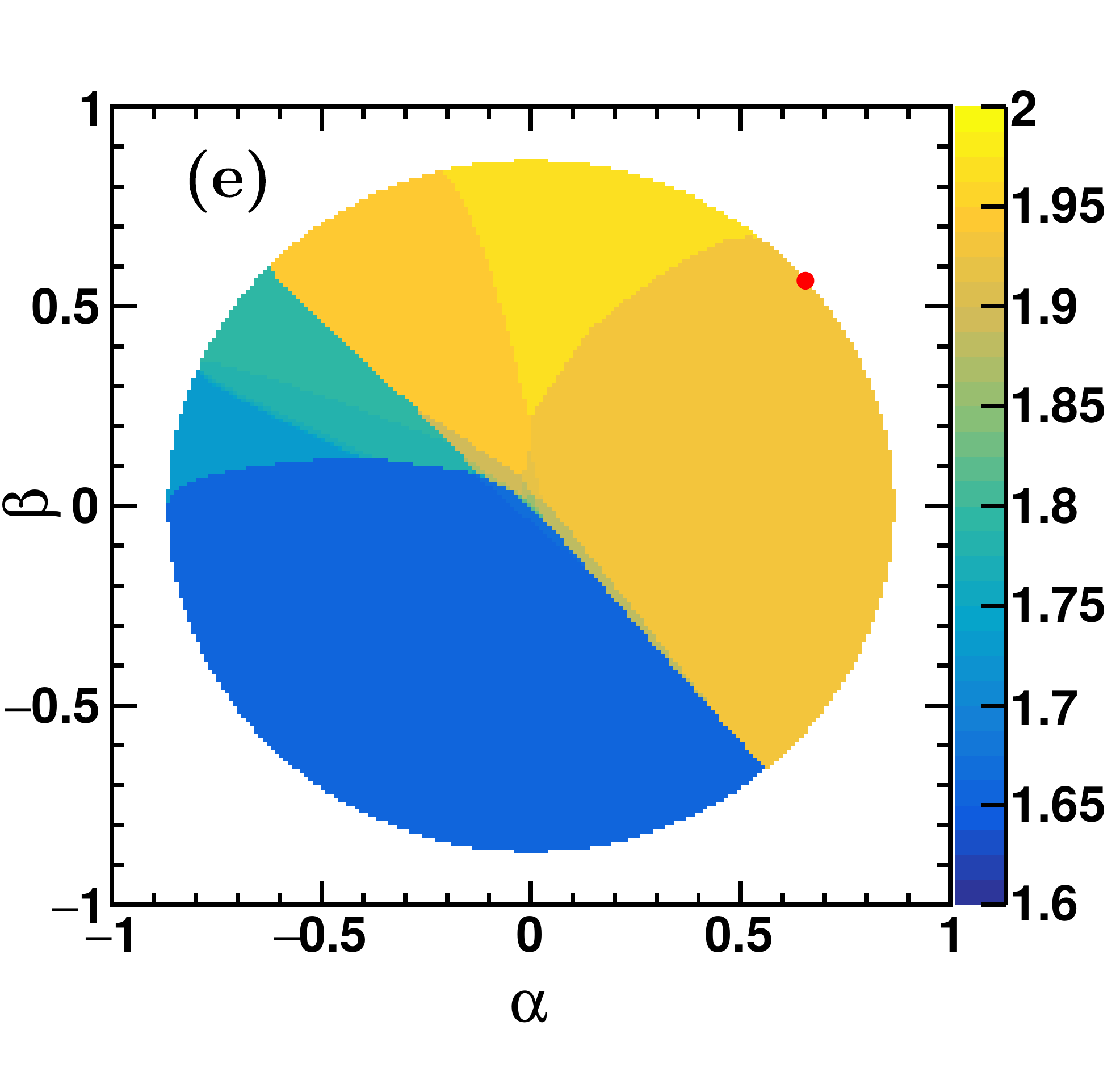} } & 
   \resizebox{65mm}{!}{ \includegraphics{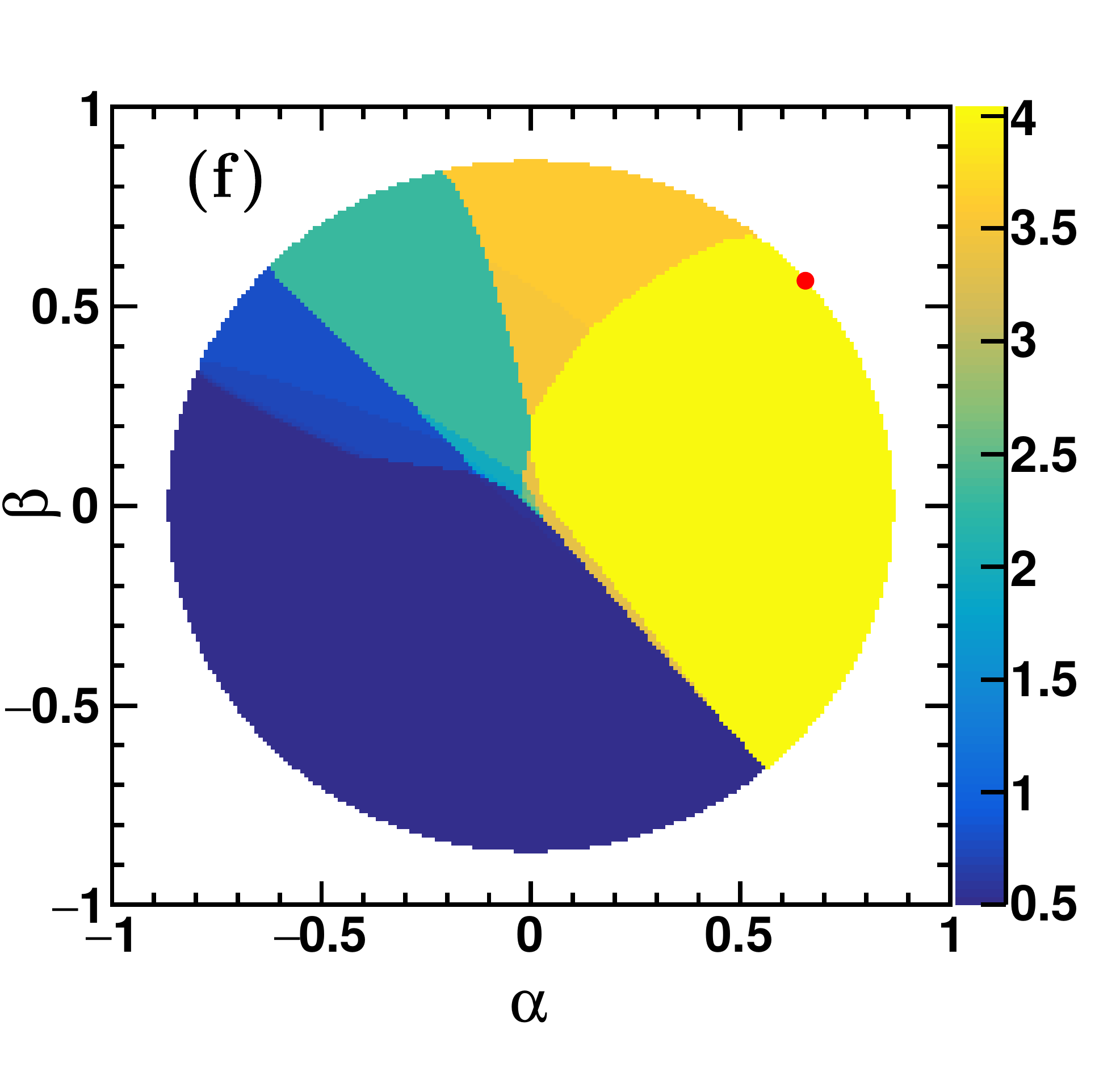} }
   \end{tabular}

  \caption{\label{fig:ana} (Color online) Results of the multivariate data set optimization for the studied case of $^8_\Lambda$Be as a function of the target thickness, beam energy, beam intensity, hypernuclear yield, exotic beam, primary beam, and target species. Figures (a) to (f) represent the evolution of the variables, the primary beam and target species, the thickness in centimeters, the selected exotic beam species and its kinetic energy in $A$GeV, and the $^8_\Lambda$Be yield per second, respectively. Each red dot represents the position of the overall maximum within the $\alpha$-$\beta$ space.}
         
\end{figure*}

A search for the maximum of the cost function is performed with fixed weight coefficients $\alpha$ and $\beta$ to determine the optimal parameter set: \[\underset{C,E,T,I}{\arg\!\max}\{F_{\alpha,\beta}(C,E,T,I)\}.\] The evolution of the obtained optimal variables according to those weight coefficients can then be investigated. First, in Fig. \ref{fig:ana_cost} the distribution of the maximum of the cost function as a function of the weights $\alpha$ and $\beta$ is presented. Additionally Fig. \ref{fig:ana} shows the different results of the optimization as a function of $\alpha$ and $\beta$.  Each value of this distribution is a possible optimal condition set. To determine the best optimal condition that should be considered, a maximax criterion was exploited.  This criterion is defined as follows: 
\begin{equation}
\label{eq:maximax}
\underset{\alpha,\beta}{\arg\!\max}\ \underset{C,E,T,I}{\max} \{F_{\alpha,\beta}(C,E,T,I)\}.
\end{equation}

The evolutions of the variable set according to $\alpha$ and $\beta$ are presented Fig. \ref{fig:ana}. The optimal isotope species of the primary beam and target are shown in Figs. \ref{fig:ana}(a) and \ref{fig:ana}(b), while the target thickness in centimeters is depicted in Fig. \ref{fig:ana}(c). The secondary beam that should be selected for the optimal production of $^8_\Lambda$Be and its resulting kinetic energy for each ($\alpha$,~$\beta$) are shown in Figs. \ref{fig:ana}(d) and \ref{fig:ana}(e) respectively. The yield per second of $^8_\Lambda$Be, produced by the collision of the optimal secondary beam and a 4-centimeter $^{12}$C target, is finally shown in Fig. \ref{fig:ana}(f) as a function of $\alpha$ and $\beta$. The evolution of the weight parameters $\alpha$ and $\beta$ within $[-\sqrt{3/4},~\sqrt{3/4}]$ represents the aim of minimizing or maximizing the influence of the variables $\mathit{Cs}$ and $\mathit{E}$ in the cost function. For instance, when $\alpha$ is negative the overall goal is to minimize the hypernuclear yield or when $\beta$ is negative the intent is to minimize the kinetic energy of the optimal secondary beam. It is useful to calculate the optimal conditions between the limits of the weights since another decision criterion can be considered instead of the maximax approach of Eq. \ref{eq:maximax}.

\begin{table*}[htb]
  \caption{Summary of the results from the optimization procedure. All $\Lambda$-hypernuclei up to carbon isotopes were considered, and for each one, the optimal experimental conditions are reported: the reaction necessary to produce the exotic beam, the target thickness, the exotic beam selected to produce the hypernuclei of interest on a 4-centimeter $^{12}$C target, the exotic beam kinetic energy and the intensity, and the resulting hypernuclear yield.}
  \label{tab:allRes}

  \begin{center}
  \begin{ruledtabular}
\begin{minipage}[t]{0.45\linewidth}
\begin{tabular}{ccccccc}
       & Reaction & Target & 2$^{nd}$ beam & E$_k$ & $I$ & Yield \\
       &  & (cm) &  & ($A$GeV) & (10$^{6}$/s)& (/s) \\ \hline
$^{8}_\Lambda$C & $^{14}$N+$^{9}$Be & 5.5 & $^{12}$N & 1.94 & 5.1 & 0.2 \\ 
$^{9}_\Lambda$C & $^{14}$N+$^{9}$Be & 5.5 & $^{12}$N & 1.94 & 5.1 & 0.8 \\ 
$^{10}_\Lambda$C & $^{14}$N+$^{9}$Be & 5.5 & $^{12}$N & 1.94 & 5.1 & 1.5 \\ 
$^{11}_\Lambda$C & $^{14}$N+$^{9}$Be & 5.5 & $^{12}$N & 1.94 & 5.1 & 0.9 \\ 
$^{7}_\Lambda$B & $^{14}$N+$^{9}$Be & 5.5 & $^{12}$N & 1.94 & 5.1 & 0.7 \\
$^{8}_\Lambda$B & $^{14}$N+$^{9}$Be & 5.5 & $^{12}$N & 1.94 & 5.1 & 2.7 \\ 
$^{9}_\Lambda$B & $^{14}$N+$^{9}$Be & 5.5 & $^{12}$N & 1.94 & 5.1 & 3.5 \\ 
$^{10}_\Lambda$B & $^{14}$N+$^{9}$Be & 5.5 & $^{12}$N & 1.94 & 5.1 & 2.5 \\ 
$^{11}_\Lambda$B & $^{20}$Ne+$^{9}$Be & 2 & $^{17}$F & 1.97 & 5.7 & 1.2 \\
$^{5}_\Lambda$Be & $^{14}$N+$^{9}$Be & 5.5 & $^{12}$N & 1.94 & 5.1 & 0.6 \\ 
$^{6}_\Lambda$Be & $^{14}$N+$^{9}$Be & 5.5 & $^{12}$N & 1.94 & 5.1 & 1.9 \\ 
$^{7}_\Lambda$Be & $^{14}$N+$^{9}$Be & 5.5 & $^{12}$N & 1.94 & 5.1 & 3.9 \\
$^{8}_\Lambda$Be & $^{14}$N+$^{9}$Be & 5.5 & $^{12}$N & 1.94 & 5.1 & 4.0 \\ 
$^{9}_\Lambda$Be & \multicolumn{2}{c}{stable beam} & $^{16}$O & 2. & 10. & 4.4 \\
$^{10}_\Lambda$Be & \multicolumn{2}{c}{stable beam}  & $^{14}$N & 2. & 10 & 3.1 \\ 
$^{11}_\Lambda$Be & $^{23}$Na+$^{11}$B & 15.5 & $^{12}$B & 1.79 & 1.2 & 0.6 \\ 
$^{4}_\Lambda$Li & $^{20}$Ne+$^{9}$Be & 2 & $^{17}$F & 1.97 & 5.7 & 1.1 \\ 
$^{5}_\Lambda$Li & $^{12}$C+$^{9}$Be & 6 & $^{10}$C & 1.94 & 5.1 & 2.5 \\ 
$^{6}_\Lambda$Li & $^{14}$N+$^{9}$Be & 5.5 & $^{12}$N & 1.94 & 5.1 & 4.3 \\ 
$^{7}_\Lambda$Li & \multicolumn{2}{c}{stable beam} & $^{14}$N & 2. & 10. & 5.2 \\ 

\end{tabular}
\end{minipage}
\hspace{0.5cm}
\begin{minipage}[t]{0.45\linewidth}

\begin{tabular}{ccccccc}

       & Reaction & Target & 2$^{nd}$ beam & E$_k$ & $I$ & Yield \\
       &  & (cm) &  & ($A$GeV) & (10$^{6}$/s)& (/s) \\ \hline

$^{8}_\Lambda$Li & $^{20}$Ne+$^{9}$Be & 2 & $^{17}$F & 1.97 & 5.7 & 3.7 \\ 
$^{9}_\Lambda$Li & $^{16}$O+$^{9}$Be & 5.5 & $^{14}$O & 1.93 & 5.5 & 2.2 \\ 
$^{10}_\Lambda$Li & $^{23}$Na+$^{11}$B & 15.5 & $^{12}$B & 1.79 & 11.5 & 1.1 \\ 
$^{11}_\Lambda$Li & $^{23}$Na+$^{11}$B & 15.5 & $^{12}$B & 1.79 & 11.5 & 0.12 \\ 
$^{3}_\Lambda$He & $^{14}$N+$^{9}$Be & 5.5 & $^{12}$N & 1.94 & 5.1 & 1.8 \\
$^{4}_\Lambda$He & \multicolumn{2}{c}{stable beam} & $^{14}$N & 2. & 10. & 4.1 \\ 
$^{5}_\Lambda$He & \multicolumn{2}{c}{stable beam} & $^{20}$Ne & 2.0 & 10. & 5.2 \\ 
$^{6}_\Lambda$He & \multicolumn{2}{c}{stable beam} & $^{12}$C & 2. & 10. & 4.8 \\
$^{7}_\Lambda$He & $^{20}$Ne+$^{9}$Be & 2 & $^{17}$F & 1.97 & 5.7 & 2.9 \\ 
$^{8}_\Lambda$He & $^{20}$Ne+$^{9}$Be & 2 & $^{17}$F & 1.97 & 5.7 & 1.4 \\ 
$^{9}_\Lambda$He & $^{23}$Na+$^{11}$B & 15.5 & $^{12}$B & 1.79 & 11.5 & 0.8 \\ 
$^{3}_\Lambda$H & \multicolumn{2}{c}{stable beam} & $^{16}$O & 2. & 10. & 5.1 \\ 
$^{4}_\Lambda$H & \multicolumn{2}{c}{stable beam} & $^{20}$Ne & 2. & 10 & 4.5 \\ 
$^{5}_\Lambda$H & \multicolumn{2}{c}{stable beam} & $^{14}$N & 2. & 10. & 3.1 \\ 
$^{6}_\Lambda$H & $^{20}$Ne+$^{9}$Be & 2 & $^{17}$F & 1.97 & 5.7 & 1.5 \\ 
$^{7}_\Lambda$H & $^{20}$Ne+$^{9}$Be & 2 & $^{17}$F & 1.97 & 5.7 & 0.5 \\ 
$^{8}_\Lambda$H & $^{23}$Ne+$^{9}$Be & 15.5 & $^{12}$B & 1.79 & 11.5 & 0.3 \\ 
$^{3}_\Lambda$n & $^{20}$Ne+$^{9}$Be & 2 & $^{17}$F & 1.97 & 5.7 & 2.1 \\
$^{4}_\Lambda$n & $^{20}$Ne+$^{9}$Be & 2 & $^{17}$F & 1.97 & 5.7 & 1.0
   \end{tabular}
   \end{minipage}
   \end{ruledtabular}
  \end{center}
\end{table*}

In the case of $^8_\Lambda$Be, shown in Fig. \ref{fig:ana}, the systematic procedure gives the following set of experimental parameters: with a primary beam of $^{14}$N impinged on a 5.5-centimeter $^9$Be target, an exotic beam of $^{12}$N should be selected and transported to bombard a $^{12}$C target. The intensity of the $^{12}$N secondary beam is about $5.1 \times 10^6$ ions per second with a primary beam of $5 \times 10^9$ ions per second, which is within the expected intensity that the new FAIR facility will provide at the entrance of the Super-FRS. Subsequently, under those conditions the $^8_\Lambda$Be yield is about 4.0 hypernuclei produced per second for a 4-centimeter secondary target.

After reviewing the details for the $^8_\Lambda$Be case, one can proceed similarly for other hypernuclei. Table \ref{tab:allRes} gathers the results of the optimization for $\Lambda$-hypernuclei up to Carbon hypernuclei. The reaction necessary for the optimal exotic beam production is reported with its target thickness. The selected exotic beam is then mentioned with its optimal kinetic energy and intensity. Finally the yield per second of the hypernucleus of interest is given for a production on a 4-centimeter $^{12}$C target. Several cases in which a stable beam provides a higher hypernuclear yield can be noted in Table \ref{tab:allRes}. Those optimal experimental conditions will be useful for conceiving the future hypernuclear experiments within the Super-FRS of FAIR facility.

\section{Conclusion} 

A general procedure was developed in order to determine the optimal experimental conditions for the production of exotic hypernuclei within the Super-FRS fragment separator of the new FAIR facility, in which future hypernuclear spectroscopy experiments will take place. This optimization process includes the results from several theoretical models. The production of hypernuclei and of exotic beams from the fragmentation of the primary beam on the production target were estimated. The procedure also includes Monte Carlo simulations of the beam transportation to estimate the experimental efficiency of the separator. Those efficiencies correspond to the transport from the secondary beam production site to the experimental area where both the hypernuclei production and spectroscopy will take place. The optimization procedure combined those different results and tried to provide the best conditions for any considered hypernucleus. The best experimental requirements were obtained by the maximization of the cost function. The optimal conditions of a $^8_\Lambda$Be hypernucleus were determined. The use of a secondary beam of $^{12}$N from the fragmentation of a $^{14}$N primary beam on a $^9$Be target was found to be optimal. Around four $^8_\Lambda$Be would be produced per second on a 4-centimeter $^{12}$C target, with an estimated $^8_\Lambda$Be production cross section of 11 $\upmu$b. Under those conditions, and considering the efficiency of the previous experiment \cite{rappold_hypernuclear_2013}, about $345 \times 10^3$ hypernuclei per day are expected. Afterwards, the design of the experimental apparatus, consisting of the detector setups and data acquisition, will provide the estimated hypernuclear count rate in the recorded data. In addition, the cases of all hypernuclei up to carbon hypernuclei were optimized and reported. This information will be valuable for further experiments on proton-rich or neutron-rich hypernuclei. Those results were achieved by a particular quasi-convex combination of the variables that were optimized. There are other possibilities to be explored, especially depending on the weight definition in the cost functions. Furthermore the maximax criterion was used and different criteria can also be applied to tune the weights. Additional considerations could provide new perspectives on the data set.

\section{Acknowledgments}

This work has been supported by the HypHI project funded by the Helmholtz association as Helmholtz-University Young Investigators Group VH-NG-239 at GSI, and the German Research Foundation (DFG) under contract number SA 1696/1-1 and EU FP7 HadronPhysics-2 SPHERE. This work has also been supported by the co-funded program of the University of Castilla-La Mancha ``Ayudas para estancias de investigadores invitados en la UCLM para el a\~no 2015" and FEDER 2014-2020. The second author has been sponsored by Ministerio de Economía y Competitividad and fondos FEDER MTM2013-47879-C2-1-P. A part of this work was carried out on the HIMSTER high performance computing infrastructure provided by the Helmholtz-Institute Mainz. We would like to thank to A. Botvina,  H. Geissel, T. Saito and C. Scheidenberger for the involved discussions.

\FloatBarrier

\bibliography{MyLibrary}

\end{document}